\newcommand{\GG}[1]{\textcolor{black}{#1}}
\newcommand{\G}[1]{\textcolor{black}{#1}}

\documentclass[]{aa}
\usepackage{graphicx,url,twoopt}
\usepackage[varg]{txfonts}
\usepackage{subfigure} 
\usepackage{enumerate}

\bibpunct{(}{)}{;}{a}{}{,} 
\usepackage[colorlinks,
            linkcolor=red,
            anchorcolor=blue,
            citecolor=blue]{hyperref}
\graphicspath{{./}{figures/}}

\begin{document}

\title{The binary fraction of Blue Horizontal-branch (BHB) Stars}
\author{{Yanjun Guo}\inst{1,2},
         Kun Chen\inst{1,2},
         Zhenwei Li\inst{1,2},
         Jie Ju\inst{3,4},
         Chao Liu\inst{5,6},
         Xiangxiang Xue\inst{5,6},
         Matti Dorsch\inst{7},
         Zhanwen Han\inst{1,2}
          \and
          XueFei Chen\inst{1,2}
          }
\institute{
            $^1$ Yunnan observatories, Chinese Academy of Sciences, P.O. Box 110, Kunming, 650011, China; cxf@ynao.ac.cn\\
            $^2$ International Centre of Supernovae, Yunnan Key Laboratory, Kunming 650216, China; guoyanjun@ynao.ac.cn\\
            $^3$ Department of Physics, Hebei Normal University, Shijiazhuang 050024, People's Republic of China\\
            $^4$ School of Sciences, Hebei University of Science and Technology, Shijiazhuang 050018, China\\
            $^5$ CAS Key Laboratory of Optical Astronomy, National Astronomical Observatories, Chinese Academy of Sciences, Beijing, 100101, People’s Republic of China\\  
            $^6$ Institute for Frontiers in Astronomy and Astrophysics, Beijing Normal University, Beijing 102206, China\\
            $^7$ Institut für Physik und Astronomie, Universität Potsdam, Haus 28, 
            Karl-Liebknecht-Str. 24/25, 14476 Potsdam, Germany
             }

\abstract
   {Blue horizontal-branch (BHB) stars are old, low-mass, metal-poor stars that serve as important tracers of the Galactic halo structure, kinematics, and evolution.Understanding their binary properties provides key insights into their formation channels and the role of binary interactions in the evolution of horizontal branch stars.
}
    {We intend to investigate the intrinsic binary fraction $f_{\rm b}^{\rm in}$ of BHB stars and its dependencies on metallicity, kinematics, and \G{effective temperature}.}
    {We collect \GG{299} BHB stars from LAMOST with multiple radial velocity (RV) measurements and classify the sample into halo-like and disk-like BHBs based on their kinematics and metallicity, as well as into bluer and redder BHBs based on their \G{effective temperature}. We then investigate the observed binary fraction for each group based on the radial velocity variations and apply a set of Monte Carlo simulations, assuming distributions of $f(P) \propto P^\pi$ and $f(q) \propto q^\kappa$, to correct the observed binary fraction for observational biases and derive the intrinsic binary fraction.
    }
    {The observed binary fraction of BHB stars is \G{$18\%\pm2\%$} for cases with $n \geq 2$ and \G{$21\%\pm3\%$} for cases with $n \geq 3$, where n represents the number of observation times. After correcting for observational biases, the intrinsic binary fraction increases to \G{$31\%\pm3\%$} for $n \geq 2$ and \G{$32\%\pm3\%$} for $n \geq 3$. 
    A clear contrast is observed between halo-like and disk-like BHB stars, with halo-like BHBs exhibiting a lower intrinsic binary fraction (\G{$28\%\pm3\%$} for $n \geq 2$ and \G{$29\%\pm3\%$} for $n \geq 3$) compared to disk-like BHBs (\G{$46\%\pm11\%$ and $51\%\pm11\%$}, respectively), indicating different formation pathways.
    In particular, halo-like BHB stars are more likely to have formed via a single-star evolution channel, whereas disk-like BHB stars may predominantly result from binary evolution processes.
    Additionally, we find that bluer BHB stars exhibit a \G{significantly} higher binary fraction (\G{$42\%\pm6\%$ for $n \geq 2$ and $45\%\pm6\%$} for $n \geq 3$) than redder BHB stars (\G{$24\%\pm5\%$ and $23\%\pm5\%$}, respectively), which suggests a possible link between binarity and the \G{effective temperature}, although more samples are required to confirm this.
    No correlation is found between $\pi$ ($\kappa$) and metallicity or kinematics, nor between $\pi$ ($\kappa$) and the \G{effective temperature} of BHB stars.} 
    {}


\titlerunning{The binary fraction of BHB stars} 
\authorrunning{Yanjun Guo et al.}        
\maketitle

\section{Introduction} \label{sec:intro}
Blue horizontal-branch (BHB) stars are low-mass (<1.0 $M_{\odot}$), metal-poor Population II stars that are typically found in the Galactic halo, with the majority being A-type stars \citep{1952Arp,2003Behr,2009Gray,2009Catelan}.
These stars are horizontal branch (HB) stars positioned to the blue side of RR Lyrae variables, where they undergo core helium fusion while sustaining hydrogen shell burning in their outer layers \citep{1966Searle,
2011Ruhland,2021Culpan}.
Similar to RR Lyrae stars, BHB stars are used as standard candles due to their nearly constant luminosity across a wide range of effective temperatures \citep{1995Dorman,1996Layden,2024Culpan}.
Given that BHB stars are over ten times more abundant than RR Lyrae stars, they are more commonly used to investigate the Galactic halo and serve as a preferred tracer for estimating its mass distribution \citep{1984Pier,1996Beers,2000Yook,2008Xue,2009Catelan,2011Xue}.

Some studies have investigated the formation of blue horizontal branch stars \citep{1952Arp,1968Castellani,1970Iben,1990Sweigart,2001Maxted,2011Ruhland,2016Girardi,2016Heber}.
\cite{1955Hoyle} were the first to accurately recognize that horizontal branch (HB) stars originate from low-mass red giant branch (RGB) stars, undergoing core helium burning while sustaining hydrogen shell burning \citep{1966Faulkner, 2001Moehler}. 
During the RGB phase, stars experience significant mass loss, with blue HB stars losing more mass than red HB stars, which accounts for their color differences \citep{2009Catelan}. 
High metallicity star clusters typically have a predominantly red horizontal branch, while low metallicity clusters generally exhibit a blue horizontal branch \citep{1966Faulkner}. 
Additionally, several studies suggest that BHB stars may also have a binary-related origin \citep{1994Liebert, 1995Bailyn, 1997Rich,2009Hubrig,2024LiZW}, although the specific mechanisms remain unclear due to limited research.

The binary fraction is very important for understanding the evolution of stellar populations and the formation of many objects \citep{2010Raghavan,2013Duchene,2017MoeStefano,2018XueFei,2019liuchaosmokinggun,2020Hanzhanwen,2024Chen}. 
However, observed fractions can be affected by observational biases \citep{2007Kobulnicky,2011Sana,2012SanaScience}. 
To correct for observational biases, \cite{2013sana} proposed a method using Monte Carlo simulations, which has been widely adopted in many studies to obtain the intrinsic binary fraction by correcting the observed fraction for these biases \citep{2012SanaScience, 2015Dunstall, 2021Banyard, 2021MahyC, 2021GYJfb, 2021luofeng}.
A homogeneous sample with multiple observations and reliable radial velocity (RV) measurements would yield more reliable results \citep{2021luofeng,2022guo886}. 

Current spectroscopic surveys, including the Sloan Digital Sky Survey (SDSS) \citep{2000Yook} and the Large Sky Area Multi-Object Fiber Spectroscopic Telescope (LAMOST) \citep{2012CuiXiangQun,2020LiuChao}, offer valuable opportunities to study stellar statistical properties with huge homogeneous samples.
\cite{2011Xue} selected 4,985 BHB stars from SDSS DR8 using color cuts and criteria derived from Balmer-line profiles. 
\cite{2024JuBHB} cataloged 5,355 BHB stars using low-resolution spectra from LAMOST.
The sample of \cite{2024JuBHB} includes multiple observations and reliable radial velocity (RV) measurements after cross-matching with the RV catalog from LAMOST MRS \citep{2021zhangboRV}.
Given the potential binary-related origin of blue horizontal branch (BHB) stars and the observed correlations between their metallicities, kinematics, and horizontal branch morphology, in this work we aim to investigate the relationship between the binary fractions of BHB stars, their metallicities and kinematics, and their positions on the Hertzsprung-Russell (HR) diagram, applying the method of \citep{2013sana} based on the sample from \cite{2024JuBHB}.

The structure of the paper is as follows. 
We introduce the LAMOST data in Section~\ref{sec:LAMOST DATA}. 
In Section \ref{sec:method}, we describe the sample grouping and criteria for the binary.
A brief description of the Monte Carlo method to correct for observational biases is provided in Section~\ref{sec: Correction for observational biases}. 
The results are presented in Sec.~\ref{sec: Results and Discussion}. 
Finally, we summarize our conclusions in Sec.~\ref{sec:Summary}.

\section{Data} \label{sec:LAMOST DATA}
The Large Sky Area Multi-Object Fiber Spectroscopic Telescope (LAMOST), located at the Xinglong Station of the National Astronomical Observatories in China, is a 4-meter quasi-meridian reflecting Schmidt telescope. 
Its innovative design allows for a wide field of view, with a focal plane spanning 5 degrees and accommodating up to 4,000 fibers simultaneously \citep{2012CuiXiangQun,2012ZhaoGang,2012DengLiCai}.
Since 2012, LAMOST has been conducting a Low-Resolution Spectroscopic Survey (LRS), characterized by a resolving power of $R\sim 1800$ and a wavelength range of  $3690 \sim 9100$~\AA\ \citep{2012DengLiCai}.
In October 2018, LAMOST initiated the Medium Resolution Survey (MRS) with a resolution of $R\sim 7500$ to collect multi-epoch observations. 
The MRS spectra obtained with the blue arm span a wavelength range from $495$ to $535$ nm, while the red arm covers a wavelength range from $630$ to $680$ nm \citep{2020LiuChao}.

By measuring the equivalent widths of multiple absorption line profiles and spectral features, \citet{2024JuBHB} identified 5,355 BHB stars from LAMOST low resolution spectra.
In this work, we utilized the BHB stars and their atmospheric parameters identified by \citet{2024JuBHB} from LAMOST DR5 and cross-matched these stars with the updated RV catalog from LAMOST DR10 based on medium-resolution spectra \citep{2021zhangboRV} to analyze their binarity.

\section{Sample selection and binary criteria}\label{sec:method}
\subsection{Sample selection}\label{sec: Classify}
Since we aim to investigate the binary properties of BHB stars, our sample selection requires each star to have at least two observations (See Sec.~\ref{sec: Obsfb}). 
Starting from an initial sample of 5,355 BHB stars, we obtained a subset of 345 sources with more than twice reliable RV measurements (signal-to-noise ratio $\geq$ 20) and robust atmospheric parameters.

\begin{figure}
	\centering
	\includegraphics[scale=0.5]{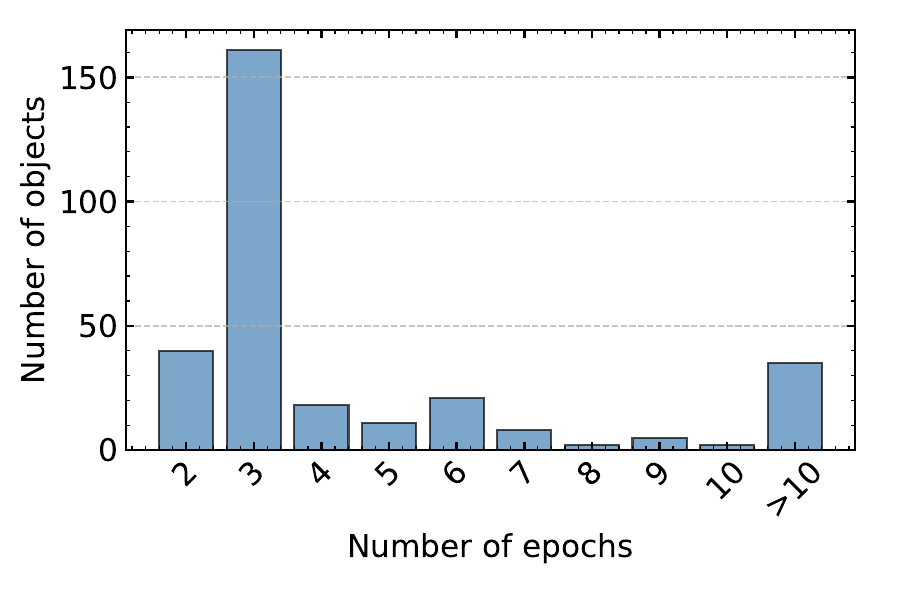}
    \caption{The distribution of the number of observational epochs for the \GG{299} BHB stars in our sample. All stars with more than ten observations are grouped into the “>10” bin.}
    \label{fig: Obstime}
\end{figure}

\G{
BHB stars are particularly susceptible to contamination from main-sequence (MS) stars.
In \cite{2024JuBHB}, the sample was filtered using equivalent width cuts on $H_{\gamma}$ and G4300, along with Balmer-line profile cuts, to remove MS contaminants. However, because this method relies on automated selection across a large sample without visual inspection, some residual contamination may remain.
Since MS contamination impacts the BHB samples differently depending on their characteristics, the detailed removal of MS contaminants is described in Sect.~\ref{sec: group} after classification.
After removal, we ultimately obtained a subset of \GG{299} sources} 
and the distribution of observation frequencies in our sample is shown in Figure \ref{fig: Obstime}.
We find that 8\G{7}\% of the sources have more than three observations, while only a few stars have a higher number of epochs.
We present the information for each star in Table \ref{tab:RV Catalogs}.

According to \citet{2013sana,2021luofeng,2021GYJfb,2022guo886}, both sample size and observation frequency significantly influence the uncertainty of the binary fraction. 
Therefore, we need to ensure sufficient observation frequency while maintaining an adequate sample size for statistical reliability.
To assess the impact of observation frequency, we tested results for samples with more than 2, 3, 4, 5, and 6 observations. 
However, we found that for 4, 5, and 6 observations, the sample size became too small for reliable statistical analysis. 
Therefore, our further analysis and main conclusions are based on the results for stars with at least 2 and 3 observations.

\begin{table*}
\caption{\label{tab:RV Catalogs}The RV catalogs.}
\centering
\begin{tabular}{lcccccccccl}
 \hline \hline
 RA    & DEC    &  MJD   & SNR & $T_\mathrm{eff}$   &  $\log{g}$  & $[M/H]$  &     $RV$      & $\sigma$\\  (deg) & (deg)  & (day)   &     &      (K)          &  (dex)   & (dex)  & ($\ {\rm km\ s^{-1}}$) & $\ {\rm km\ s^{-1}}$)\\ \hline
223.29825  &6.97886  & 59618.58  &25 &7922 &4.2  &-0.9   &-165.11    &2.60\\
223.29825  &6.97886  & 59618.56   &24 &7922 &4.2  &-0.9  &-162.08    &2.67\\
234.54288  &38.63063  & 58887.57   &23  &11033 &3.8  &-1.2   &-151.44   &2.78\\
234.54288  &38.63063  & 58887.56    &22  &11033 &3.8 &-1.2  &-153.06    &2.81\\
190.92388  &32.35772 & 58507.51    &21  &8981 &3.8  &-1.5  &148.78   &2.51\\
190.92388  &32.35772  & 58953.28   &48  &8981 &3.8  &-1.5  &146.99    &1.78\\
\hline  \hline
\end{tabular}
\tablefoot{This table is available in its entirety in machine-readable form. The first six entires are shown here for guidance regarding its format and content.}
\end{table*}

\subsection{Grouping of the sample}\label{sec: group}
\begin{figure}
	\centering
	\includegraphics[scale=0.5]{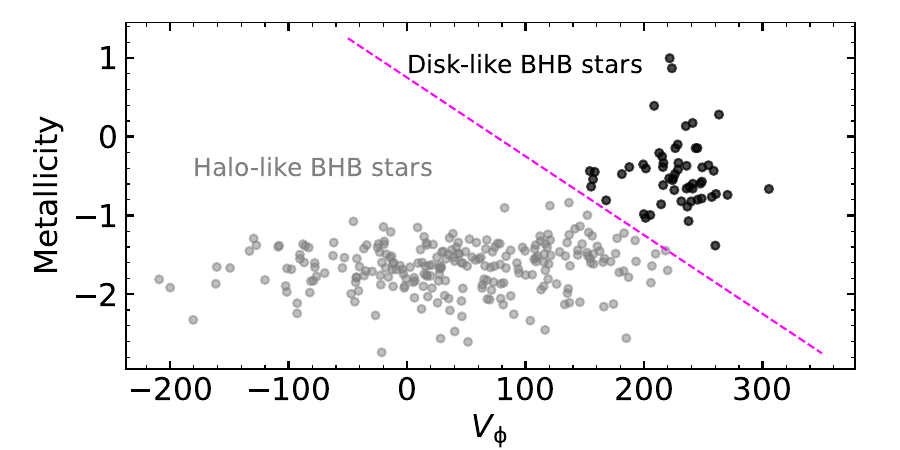}
    \caption{Classification of BHB stars based on kinematics and metallicity. The magenta dashed line represents the boundary used to separate halo-like (gray) and disk-like (black) BHB stars in the plane of azimuthal velocity and metallicity.}
    \label{fig: fehV}
\end{figure} 

\begin{figure}
	\centering
	\includegraphics[scale=0.5]{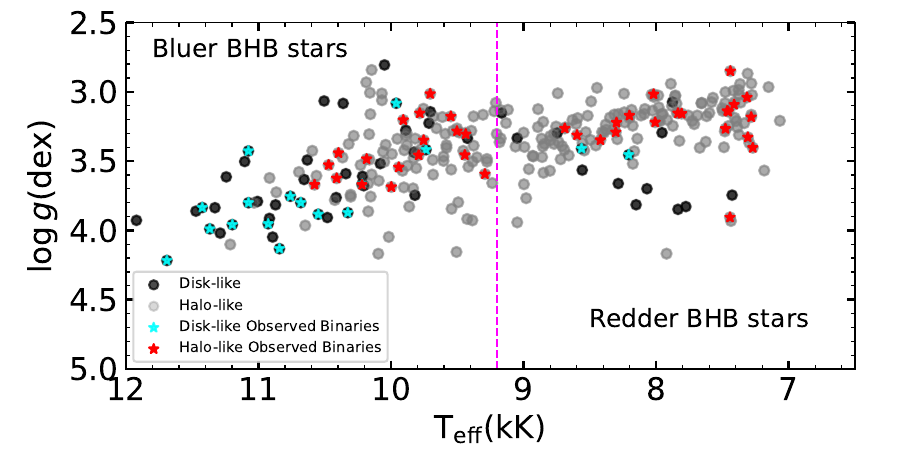}
    \caption{Classification of BHB stars based on their positions on the \G{Kiel diagram}. The magenta dashed line marks the boundary distinguishing redder BHB stars and bluer BHB stars in the plane of effective temperature and surface gravity. Gray dots represent halo-like BHB stars, while black dots denote disk-like BHB stars. Halo-like observed binaries (red stars) and disk-like observed binaries (cyan stars) are highlighted accordingly.}
    \label{fig: TeffG}
\end{figure} 
According to the age-metallicity relation, stars with higher metallicity generally tend to be younger, while those with lower metallicity are typically older \citep{1980Twarog,1998Carraro,2000Rocha-Pinto}.
Additionally, azimuthal velocity ($V_{\phi}$) provides insight into the Galactic component to which these stars belong, with higher $V_{\phi}$ values indicating significant rotational motion, characteristic of disk stars, and lower $V_{\phi}$ values suggesting weaker rotational support, consistent with halo stars \citep{2019tianhao,2020Belokurov,2024tianhao}. 
Figure \ref{fig: fehV} presents the relationship between azimuthal velocity ($V_{\phi}$) and metallicity (\text{[Fe/H]}) for our sample, where the data points are divided into two main groups using the magenta dashed line.
Based on these two properties, we classify stars with higher metallicity and higher azimuthal velocity as disk-like BHB stars, while those with lower metallicity and lower azimuthal velocity are classified as halo-like BHB stars.
This classification is motivated by the investigation of the correlation between metallicity, kinematics, and binary fraction.
\G{It is worth noting that the [Fe/H]–V$_{\phi}$ classification is not absolute: some stars with [Fe/H] $\sim$ –1 and high V$_{\phi}$ may still be genuine halo members, while those with [Fe/H] $\sim$ –1.5 and V$_{\phi}$ $\sim$ 200 km/s could in fact belong to the thick disk.}
To further investigate the relationship between binary fraction and \G{effective temperature}, we divided our sample into two groups based on effective temperature \G{using $T_{\rm eff}$ = 9200 K as the boundary, which approximately corresponds to the midpoint of our sample’s temperature range.} 
It is worth noting that under these two classifications, we have also ensured that the number of detectable binaries in each group is \G{comparable} (see the $N_{b}$ column in Table 3), since too few detectable binaries may lead to increased random errors.

\G{After the classification, we subsequently excluded residual MS contaminants from each group, as motivated by the reasons detailed in Section\ref{sec: Classify}.
Since MS contamination affects only the disk sample (hot MS stars are young and not expected in halo), we further cleaned the disk-like BHB sample with the following step.
\citet{2013Salgado} show that horizontal‐branch stars with $T_{\rm eff} \leq 10,\GG{500},$K inhabit $\log g \sim 3.0-4.0$ dex, whereas main‐sequence stars at similar $T_{\rm eff}$ lie at $\log g>4.0$. We therefore removed 26 redder disk‐like ($T_{\rm eff} \leq 10,\GG{500},$K) candidates with $\log g>4.0$ dex.
\GG{Additionally, recognizing that some halo-like stars may be misclassified non-halo objects, we applied the same criterion to the halo-like BHB group and further removed four stars.}
We derived the absolute $V$-band magnitude ($M_V$) for each star by calculating the absolute Gaia $G$-band magnitude ($M_G$) and applying the photometric transformation coefficients provided in the Gaia Data Release 3 documentation\footnote{\url{https://gea.esac.esa.int/archive/documentation/GDR3/Data_processing/chap_cu5pho/cu5pho_sec_photSystem/cu5pho_ssec_photRelations.html}}.
The distances and extinction values required in this process were obtained from Ju (in prep.), who constructed new BP-band absolute magnitude–(BP–RP) relations for BHB stars based on accurate Gaia distances estimated by \citet{2021Bailer-Jones}. 
These values were then derived through repeated iterations.
Because genuine red HB stars are typically brighter than MS stars, with $M_{V} \sim 0.5$–$1.0$, and the sample becomes increasingly dispersed above $M_{V} >1.3$, we excluded 16 disk‐like BHB candidates with $M_{V} > 1.3$ to minimize MS contamination.
Unfortunately, for the hotter disk‐like BHBs, the difference in $M_{V}$ between genuine BHBs and MS stars becomes too small to separate cleanly.
In addition, even after applying the spectral profile diagnostics described in \cite{2024JuBHB}, we cannot effectively remove MS contaminants from the hotter BHB sample.
Therefore, throughout Section~\ref{sec: Results and Discussion}, we highlight, for every relevant result, the limitations associated with hotter disk‐like BHB stars, which may be affected by MS contamination.
}
As shown in Figure~\ref{fig: TeffG}, a magenta \G{vertical line at $T_{\rm eff}$ = 9200 K} separates the hotter, bluer BHB stars from the cooler, redder BHB stars.
The halo-like BHB stars are depicted as gray dots, while disk-like BHB stars are represented by black dots.
Observed binaries (See sec.~\ref{sec: Obsfb}) are distinguished by color, with red stars indicating halo-like binaries and cyan stars indicating disk-like binaries.

\subsection{Criterion for the binary}\label{sec: Obsfb}
To identify binary systems in our sample, we applied the criterion from Equation 4 in \cite{2013sana}, which has been widely used in previous studies \citep{2012SanaScience,2015Dunstall,2021luofeng,2021MahyC,2021Banyard,2021GYJfb}. This criterion classifies a star as a binary if its radial velocities (RVs) satisfy the following conditions:

\begin{equation}
\centerline{ $\frac{|v_{i} - v_{j}|}{\sqrt{\sigma_{i}^2 \ + \sigma_{j}^2}}\  > \ 4$, and ${|\\ v_{i} - v_{j} \\ |}\  >  \ C$,}\label{cer:SB1}
\end{equation}
where $v_{i(j)}$ represents the RV measured from the spectrum at epoch i~(j), and $\sigma_{i(j)}$ is the corresponding uncertainty. 
To mitigate the significant RV variations caused by stars with pulsations, \cite{2013sana} introduced threshold  C  here as a filter to remove pulsating stars.
Based on the kinks observed in the RV distributions of their sample, \cite{2013sana} adopted C=20~$\ {\rm km\ s^{-1}}$ for O-type stars, 
while \cite{2015Dunstall} set C=16 $\ {\rm km\ s^{-1}}$ for B-type stars.
\G{
For our sample, the kink is weak, so it is more reasonable to adopt C based on the typical RV amplitudes of potential pulsators.
Blue horizontal branch (BHB) stars are not located in the classical instability strip, and thus pulsating stars are rare among them \citep{2010Aerts,2019GaiaCollaboration,2025GaoXinyi}. 
To date, only one such pulsator has been identified \citep{2012Ostensen}, indicating that a significant number of pulsating stars are not expected in our sample, and thus the threshold is unnecessary.
We therefore adopted C = 0 $\ {\rm km\ s^{-1}}$ for our subsequent analysis.
Additionly, we performed tests with C$= 10$ and 20 km/s to assess their impact on the intrinsic binary fractions, as discussed in Sec.~\ref{sec: Results and Discussion}. 
}

\section{Correction for observational biases}\label{sec: Correction for observational biases} 
\begin{table*}
\caption{\label{tab:sim parameter}The range of different parameters and Power Indexes used in MC simulation. 
}
\centering
\begin{tabular}{lccccl}
 \hline \hline
Parameter  & Power law  &  Parameter Range  &  Power Index &  Index Range &Step\\\hline
$P$(d)   &$f(P) \propto P^\pi$    & 1 - 1000        &$\pi$    & $-7.00$ - $-2.00$      &0.05\\
$q$      &$f(q) \propto q^\kappa$ & 0.1 - 5.0       &$\kappa$ & $-3.00$ - $\ \,2.00$         &0.05\\
$f_{\rm b}$  &-                       &-                &-        &$\, \, 0.00$ - $\, \, 1.00$      &0.01\\
  \hline  \hline
\end{tabular}
\tablefoot{$\pi$ and $\kappa$ are the power index of $P$ (orbital period) and $q$ (mass ratio), respectively.
$f_{\rm b}$ is the binary fraction for simulation.
The ranges of $P$ and $q$ are shown in Parameter Range, while the ranges of Power Index are shown in Index Range.
The last row gives the step of each Power Index.}
\end{table*}

\subsection{Monte-Carlo method}\label{sec: Monte-Carlo method} 
Based on observed binary fraction ($f_{\rm b}^{\rm obs}$), we applied several Monte-Carlo (MC) simulations, similar to those of \cite{2013sana},\cite{2015Dunstall}, \cite{2022guo886} to assess the intrinsic binary fraction ($f_{\rm b}^{\rm in}$).
To conduct the simulations, we need to generate two synthetic cumulative distribution functions (CDFs): one for the radial velocity variance ($\Delta$RV), which represents the maximum RV variation across individual exposures, and another for the minimum time interval between exposures ($\Delta$MJD).
For the orbital period $P$ and mass ratio $q$, we adopt a power-law distribution \citep{2013sana}, expressed as  $f(P) \propto P^\pi$ and $f(q) \propto q^\kappa$, using \rm log $P$  instead of (log\ $P)^\pi$  because the linear form is more sensitive to short-period binaries \citep{2021GYJfb}.
Tab.~\ref{tab:sim parameter} lists the variable ranges for the above parameters and power index. 

Generally, a two-body kinetic system in Keplerian motion is characterized by several orbital parameters: the inclination (i), the semi-major axis (a), the argument of periastron ($\omega$), the eccentricity (e), and the epoch of periastron ($\tau$). 
The inclination is randomly selected between 0 and $\pi$/2 following a $sin(i)$ probability distribution. 
The semi-major axis is related to the orbital period (P). 
Meanwhile, $\omega$ is uniformly distributed over the range 0 to 2$\pi$. 
The eccentricity is modeled by a distribution proportional to e raised to the power of $e^\eta$.
Given that the global merit function is insensitive to $\eta$ (see \cite{2013sana}), we adopt -0.5 as in \cite{2013sana}.
$\tau$ is chosen at random, with days as its unit.
Based on these assumptions, we can generate simulated radial velocity (RV) data, which then allows us to derive the cumulative distribution functions (CDFs) for both $\Delta$RV and $\Delta$MJD.

To evaluate the final results, we compare the constructed simulated CDFs with the observed ones using the Kolmogorov-Smirnov (KS) test, and the simulated fraction of detected binaries ($f_{\rm b}^{\rm sim}$) are analyzed through a binomial distribution \citep{2013sana}. 
Finally, we employ the global merit function (GMF) from \cite{2013sana} to determine the best-fitting results.
The GMF consists of the KS probabilities of $\Delta$RV and $\Delta$MJD distributions and the Binomial probability: 
\begin{equation}
\centerline{ GMF= $P_{ks}$($\Delta$RV) $\times $ $P_{ks}$($\Delta$MJD) $\times $ $B(N_{b},N,f_{\rm b}^{\rm sim}$)}
\end{equation}
where $N_{b}$ represents the number of binaries in the observations, and $N$ denotes the sample size.

In our previous work by \cite{2021GYJfb}, we revisited the dataset of 360 O-type stars from \cite{2013sana} and confirmed that its statistical properties align with those presented in \cite{2013sana}. Moreover, \cite{2021luofeng}, \cite{2021GYJfb}, and \cite{2022guo886} have successfully applied this method to LAMOST data, further demonstrating its applicability to our current study.

\section{Results and discussions}\label{sec: Results and Discussion}
\begin{figure*}
	\centering
	\includegraphics[scale=0.3]{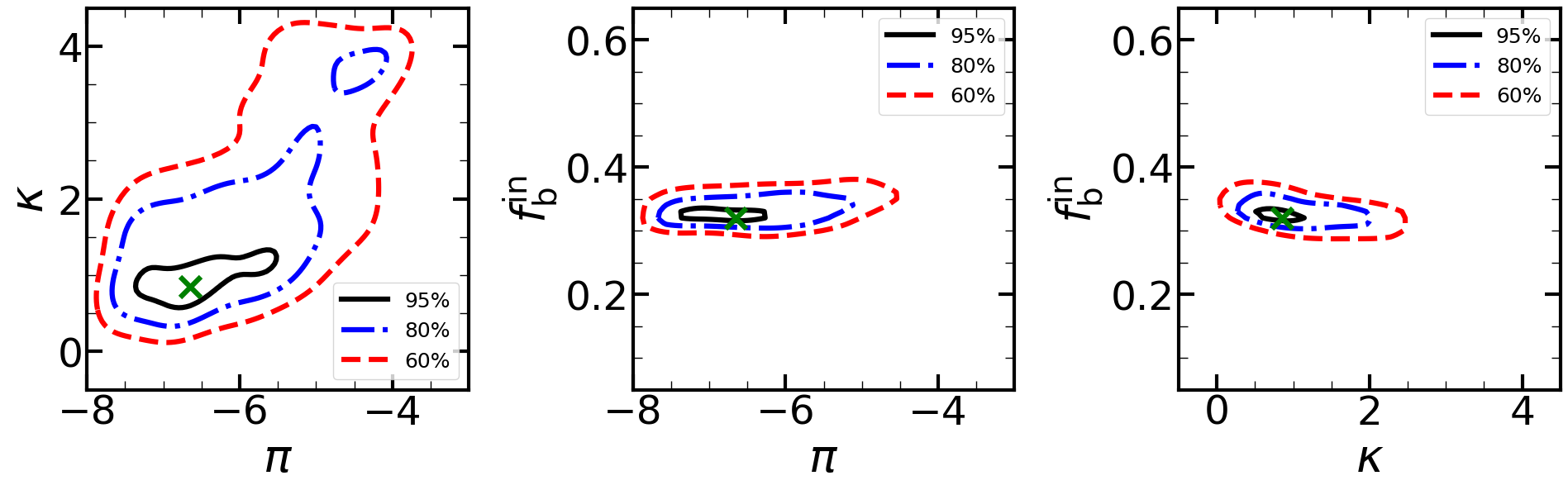}
    \caption{Projections of the GMF onto planes defined by different pairs of $\pi$, $\kappa$, and $f_{\rm b}^{\rm in}$. The green `x' marks the position of the absolute maximum. The red, blue, green, and black contours indicate regions of equal values corresponding to 60$\%$, 80$\%$, and 95$\%$ of the absolute maximum of the GMF, respectively.}
    \label{fig: GMF}
\end{figure*}

According to the criterion (\ref{cer:SB1}), the observed binary factions of our BHB stars are \G{$18\%\pm2\%$} for cases with n $\geq$ 2, and \G{$21\%\pm3\%$} for cases with n $\geq$ 3, where n represents the number of observations for each sample.
These results are summarized in Tab.~\ref{tab:fehfb}. 
The error bars of $f_{\rm b}^{\rm obs}$ were estimated through bootstrap analysis \citep{2010Raghavan,2021GYJfb}.
\G{In Tab.~\ref{tab:fehfb}, we find that none of the 55 detected binary were detected with only n = 2 epochs, so adding them to the n $\geq$ 3 sample leaves $f_{\rm b}^{\rm obs}$ essentially unchanged. 
To further quantify this, we ran Monte Carlo simulations assuming a 30\% intrinsic binary fraction.  
For n = 2,  the simulations show that only about 3 of 55 binaries are detected, demonstrating the very low detection efficiency with only two epochs.
}

After the correction for observation biased (See Sec.~\ref{sec: Correction for observational biases}), the $f_{\rm b}^{\rm in}$ of our BHB stars are \G{$31\%\pm3\%$} for cases with n $\geq$ 2, and \G{$32\%\pm3\%$} for cases with n $\geq$ 3, where the error bars are derived from the larger value between $\rm p84$ - $\rm p50$ and $\rm p50$ - $\rm p16$, corresponding to the 50th ($\rm p50$), 16th ($\rm p16$), and 84th ($\rm p84$) percentiles of the intrinsic value distribution obtained from 50 synthetic datasets \citep{2013sana,2015Dunstall}. 

\G{
To assess the impact of $C$ on the intrinsic binary fraction, we performed additional tests using $C = 10$ and $20$ km/s. 
For our BHB sample, we obtained intrinsic binary fractions of 29\% and 20\% for C =  10 and 20 km/s, respectively.
This difference arises because increasing $C$ changes the sample composition, which alters the observational cumulative distribution functions of both $\Delta$RV and $\Delta$MJD, ultimately leading to variations in the final results.
In addition, the effectiveness of the correction depends on the total number of detected binaries.
As $C$ increases, the number of detected binaries ($N_\mathrm{bin}$) decreases significantly, from 55 at $C = 0$ km/s to 19 at $C = 20$ km/s.
For smaller samples, the correction of $f_{\rm b}^{\rm in}$ is more susceptible to statistical uncertainties, thus exerting a greater influence on the final results.
}

The values of $\pi$ and $\kappa$ are consistent within the uncertainty range, with our study yielding \G{$\pi=-7.2\pm0.8$, and $-7.5\pm0.8$, and $\kappa=1.1\pm1.3$, and $0.9\pm1.3$} for n $\geq$ 2 and n $\geq$ 3, respectively.
Figure~\ref{fig: GMF} illustrates the projections of the GMF for n $\geq$ 3 as an example, with the green “x” marking the position of the absolute maximum of $f_{\rm b}^{\rm in}$, $\pi$ and $\kappa$.

\subsection{Dependency on metallicities and kinematics}\label{sec: teff}  
\begin{table*}
\caption{\label{tab:fehfb} Observed and intrinsic binary fractions for all BHB stars, as well as for halo-like, disk-like, redder, and bluer BHB subgroups, including sample size $N$ and number of observed binaries $N_{b}$.}
\centering
\begin{tabular}{lccccc}  
\hline \hline
& OBS time & N & $N_{b}$ & $f_{\rm b}^{\rm obs}$ & $f_{\rm b}^{\rm in}$ \\ 
\hline
             &2  & \GG{299} & \G{55} &\G{$18\%\pm2\%$} & \G{$31\%\pm3\%$} \\
\textbf{All} &3  & \G{263} & \G{55} &\G{$21\%\pm3\%$} & \G{$32\%\pm3\%$} \\
\hline
\hline
                   &2  & \G{247} & \G{38} &\G{$15\%\pm2\%$} & \G{$28\%\pm3\%$} \\
\textbf{Halo-like} &3  & \G{214} & \G{38} &\G{$18\%\pm3\%$} & \G{$29\%\pm3\%$} \\
\hline 
                   &2  & \G{56} & \G{17} & \G{$30\%\pm6\%$} & \G{$46\%\pm11\%$} \\
\textbf{Disk-like} &3  & \G{49} & \G{17} & \G{$35\%\pm7\%$} & \G{$51\%\pm11\%$} \\
\hline
\hline
                &2  & \G{159} & \G{22} & \G{$14\% \pm3\%$} & \G{$24\% \pm5\%$} \\
\textbf{Redder} &3  & \G{139} & \G{22} & \G{$16\% \pm3\%$} & \G{$23\% \pm5\%$} \\
\hline
               &2  & \G{144} & \G{33} & \G{$23\% \pm4\%$} & \G{$42\% \pm6\%$} \\
\textbf{Bluer} &3  & \G{124} & \G{33} & \G{$27\% \pm4\%$} & \G{$45\% \pm6\%$} \\
\hline
\end{tabular}
\end{table*}

The observed binary fraction is \G{$15\%\pm2\%$ and $18\%\pm3\%$} for halo-like BHB stars and \G{$30\%\pm6\%$ and $35\%\pm7\%$} for disk-like BHB stars for  n $\geq$ 2 and n $\geq$ 3, respectively. 
Our results reveal a difference in the $f_{\rm b}^{\rm in}$ between halo-like and disk-like BHB stars. 
For cases with n $\geq$ 2, the $f_{\rm b}^{\rm in}$ is \G{$28\%\pm3\%$} for halo-like BHB stars and \G{$46\%\pm11\%$} for disk-like BHB stars, while for n $\geq$ 3,the $f_{\rm b}^{\rm in}$ is \G{$29\%\pm3\%$} for halo-like BHB stars and \G{$51\%\pm11\%$}. 
These results are summarized in Tab.~\ref{tab:fehfb}. 

The low binary fraction of halo-like BHB stars may be explained as follows.
Horizontal branch (HB) stars are generally formed from low-mass stars when they remove some envelope and ignite central helium in a degenerate core (He flashes; \citealt{1990Sweigart,1994Liebert,1995Dorman}).
For halo-like BHB stars, 
the progenitors (before He flash) are older (and less massive) 
and have lower metallicity than that of disk-like BHB stars \citep{1990Lee,1994Lee}. 
For stars with low mass and low metallicity, 
helium ignition in single-star evolution results in relatively high temperatures, directly placing them near the BHB region, 
with many likely originating from disrupted globular clusters \citep{2011Xue}.
The low binary fraction (\G{28-29\%}) of halo-like BHB stars aligns with this scenario and indicates binary interaction has a little contribution to halo-like BHB stars.
However, this fraction is not particularly low.

\citet{2019Moe} report that the intrinsic binary fraction of metal-poor giants branch is approximately 35\%–55\% across the metallicity range $-3.5 < \mathrm{[Fe/H]} < -1.0$,
\GG{while metal-rich giants with $\mathrm{[Fe/H]} > -1.0$ exhibit lower binary fractions of about 20\%–35\%.} 
Since BHB stars generally originate from red giant branch (RGB) stars, \GG{comparing the binary fractions of these progenitor populations with those of BHB stars provides key insight into the role of binarity in BHB formation \citep{1955Hoyle,1976Sweigart,1993Dorman}.
If binary interaction plays a dominant role in the formation of BHB stars, one would expect the BHB binary fraction to be significantly higher than that of their RGB progenitors. Conversely, comparable binary fractions would suggest that binary interaction may not be a key factor.
The comparable binary fractions of metal-poor RGB stars (35\%–55\%) and our halo-like BHBs (29\%), together with the even higher binary fraction of metal-poor low-mass MS stars (54\%; \citealt{2019Moe}) and intermediate-mass MS stars ($\sim$50\%; \citealt{2017Gao}),}
suggest that most binary systems may be preserved during the evolution from the main sequence through the RGB to the BHB stage, or alternatively, that BHB binaries may originate from different evolutionary channels.

For metal-rich stars evolving via single-star evolution typically evolve into red clump (RC) stars rather than BHB stars because the lower temperature during helium burning causes them to directly settle on the RC \citep{2016Girardi,2025Moehler}. 
Therefore, for metal-rich stars to appear on the BHB, they may have much thinner envelope—potentially achieved either through enhanced mass loss via stellar winds or through binary interactions.
\GG{The significantly higher binary fraction of disk-like BHBs (51\%) compared to both metal-rich RGB stars (20\%–35\%) and metal-rich low-mass MS stars (17\%) strongly supports a binary-related formation scenario for these stars.
Although metal-rich intermediate-mass MS stars exhibit binary fractions of 30\%–60\% \citealt{2017Gao,2019Moe,2022guo886}, comparable to our disk-like BHBs, suggesting that binarity may not play a significant role in this case.
However, due to the IMF, their contribution is small and may have little effect on the disk-like BHB binary fraction.
}
Several studies have proposed a possible binary-related origin for BHB stars \citep{1994Liebert,1995Bailyn,1997Rich,2009Hebersdb}.
Similar to the formation of Extreme Horizontal Branch (EHB) stars, stable Roche lobe overflow (RLOF), common envelope (CE) ejection and merger channel are plausible pathways for removing the hydrogen envelope \citep{2001Maxted,2002Han,2008Han}. 
Additionally, halo-like BHB stars are generally older than disk-like BHB stars and exhibit a lower binary fraction, which may be partly attributed to age differences.
According to the simulations of \cite{2008Han}, as stellar populations age (i.e., transitioning from young, disk-like systems to old, halo-like systems), the dominant formation channel for EHB stars shifts from stable mass transfer to mergers, resulting in a lower binary fraction among older stars.
This trend may also be relevant for BHB stars, suggesting that similar binary interaction processes influence the formation and binary properties of both EHB and BHB stars.
While the dominant mechanism remains uncertain, the higher binary fraction in metal-rich BHB stars (\G{45-48\%}) compared to halo-like BHBs supports the idea that binary interactions are a key factor in their formation.
\G{In addition, we find that the disk-like group contains more hotter stars, so the higher binary fraction may also result from this temperature distribution (see Fig.~\ref{fig: Teff_hist} and the discussion in Section.~\ref{sec: teff}).
Notably, because we cannot entirely rule out residual MS contamination in the disk-like sample, the reported binary fraction may be slightly overestimated.}\\

The values of $\pi$ and $\kappa$ are consistent within the uncertainty range. For halo-like BHB stars, our study yields \G{$\pi = -5.8 \pm 1.4$ and $-5.4 \pm 1.4$, and $\kappa = 0.7 \pm 1.6$ and $0.9 \pm 1.6$} for $n \geq 2$ and $n \geq 3$, respectively.  
Similarly, for disk-like BHB stars, we obtain \G{$\pi = -6.5 \pm 0.6$ and $-6.0 \pm 0.6$, and $\kappa = 1.1 \pm 1.3$ and $3.6 \pm 1.3$} for $n \geq 2$ and $n \geq 3$, respectively.  

\subsection{Dependency on \G{Kiel diagram}}\label{sec: teff}  

\begin{figure}
	\centering
	\includegraphics[scale=0.5]{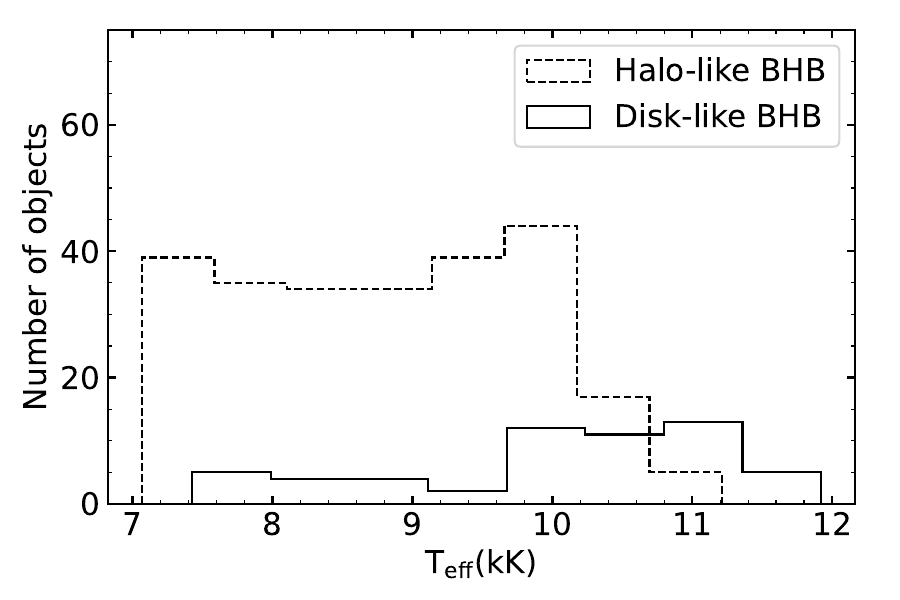}
    \caption{Histogram of the number of halo-like BHB (black dashed step) and disk-like BHB (black solid step) stars as a function of their \(T_\mathrm{eff}\) (in kK).}
    \label{fig: Teff_hist}
\end{figure}

The observed binary fraction is \G{$23\%\pm4\%$ and $27\%\pm4\%$} for bluer BHB stars and \G{$14\%\pm3\%$ and $16\%\pm3\%$} for redder BHB stars for  n $\geq$ 2 and n $\geq$ 3, respectively. 
We then examined the $f_{\rm b}^{\rm in}$ at different positions on the \G{Kiel diagram} and found that the bluer BHB sample exhibits a \G{significantly} higher $f_{\rm b}^{\rm in}$ (\G{$42\%\pm6\%$ for $n \geq 2$ and $45\%\pm6\%$} for $n \geq 3$) than the redder BHB sample (\G{$24\%\pm5\%$ for $n \geq 2$ and $23\%\pm5\%$} for $n \geq 3$).
These results are summarized in Tab.~\ref{tab:fehfb}.

Extreme Horizontal Branch stars, which are core-helium burning stars exhibiting lower masses and higher temperatures than canonical Blue Horizontal Branch stars, are strongly linked to binary interactions \citep{,2001Maxted,2002Han,2003Han,2009Hebersdb}.
The minimum binary fraction for EHB stars is inferred to be $60\% \pm 8\%$ based on \citep{2001Maxted}, while \cite{2003Han} report an intrinsic binary fraction ranging from $76\%$ to $89\%$.
Stable Roche-lobe overflow and common envelope evolution serving as the main channels for envelope stripping in red giant stars. 
These processes effectively remove the outer envelope, leaving behind a core-helium burning star with a lower mass and a higher effective temperature. 
Similarly, Blue Horizontal Branch (BHB) stars are also formed through mass loss on the red giant branch. 
In this context, the formation mechanisms of BHB stars and sdOB stars may follow a common evolutionary channel, with BHB stars experiencing just a less extreme degree of envelope removal.
High envelope removal efficiency may favor evolution into an Extreme Horizontal Branch (EHB), whereas lower efficiency may result in the star remaining as a Blue Horizontal Branch object.
The bluer a BHB star is (i.e., the closer it is to the EHB region), the more extensive the envelope stripping required. 
This may make the single-star channel increasingly unlikely and suggest that binary interactions could play a significant role in achieving such stripping.
This aligns with our finding that the bluer BHB sample exhibits a \G{significantly} higher intrinsic binary fraction (\G{45\%}) compared to the redder BHB sample (\G{23\%}) as well as observed binary fraction.
These observations suggest that the actual envelope ejection process in CE evolution may be less “clean” or efficient than idealized models predict, and this evidence can help constrain the physical parameters and efficiency of envelope stripping in CE evolution, potentially refining our understanding of binary evolution.

Fig.~\ref{fig: Teff_hist} displays the number of halo-like BHB and disk-like BHB stars as a function of effective temperature. 
\G{
The histogram shows that halo-like stars are systematically cooler, while the disk-like sample contains more hotter stars.
This distinction may leads to higher overall binary fraction observed in the disk sample because hotter BHB stars have higher binary fractions.
}
\GG{It is important to note that we cannot completely eliminate MS contamination from the hotter disk‐like BHB sample. Any residual contaminants may still lead to a modest overestimation of the binary fraction.}

In Figure~\ref{fig: TeffG}, we find that the halo-like BHB stars are relatively evenly distributed across the diagram, and their observed binaries (red stars) also do not exhibit pronounced clustering. 
In contrast, the disk-like observed binaries (cyan stars) show a tendency to cluster toward the bluer region. 
Although this may hint at a higher fraction of detectable binaries in those specific areas, the pattern should be viewed cautiously, as more data would be needed to confirm any definitive trend.

The trend of $\pi$ and $\kappa$ remains consistent within the uncertainty range across different subsets of BHB stars.  
For bluer BHB stars, our analysis yields \G{$\pi = -5.4 \pm 1.9$ and $-6.2 \pm 1.9$}, with corresponding $\kappa$ values of \G{$0.5 \pm 1.4$ and $0.4 \pm 1.4$} for $n \geq 2$ and $n \geq 3$, respectively.  
Likewise, for redder BHB stars, we derive \G{$\pi = -6.4 \pm 1.1$ and $-6.1 \pm 1.1$}, along with \G{$\kappa = 3.4 \pm 1.2$ and $3.5 \pm 1.2$} for $n \geq 2$ and $n \geq 3$.

\section{Summary}\label{sec:Summary} 
Blue horizontal-branch stars are valuable tracers of the Galactic halo, commonly used to study its structure and kinematics. Their nearly constant luminosity across a wide range of temperatures makes them reliable standard candles. Investigating their binary properties provides crucial insights into their formation pathways and the role of binary interactions in stellar evolution.

In this work, we first selected a sample of \GG{299} BHB stars with multiple observations from the catalog of \citet{2024JuBHB}. We then classified the sample into disk-like and halo-like groups based on their metallicities and kinematics, and separately into bluer and redder groups according to their \G{effective temperature}, in order to investigate their relationship with the binary fraction.  
Using the Monte Carlo Markov Chain (MCMC) framework developed by \citet{2013sana}, we corrected the observed binary fraction for any observational biases.

The intrinsic binary fraction of all \GG{299} BHB stars are \G{$31\%\pm3\%$ for $n \geq 2$ and $32\%\pm3\%$} for $n \geq 3$.  
A difference was observed in the intrinsic binary fraction between halo-like and disk-like BHB stars. Halo-like BHB stars exhibited an $f_{\rm b}^{\rm in}$ of \G{$28\%\pm3\%$ for $n \geq 2$ and $29\%\pm3\%$} for $n \geq 3$, whereas disk-like BHB stars had a much higher binary fraction of \G{$46\%\pm11\%$ and $51\%\pm11\%$} for $n \geq 2$ and $n \geq 3$, respectively, indicating that binary interactions play a more significant role in disk-like BHB stars.  
We further examined the binary fraction at different positions on the \G{Kiel diagram} and found that the bluer BHB sample exhibited a significant higher $f_{\rm b}^{\rm in}$ (\G{$42\%\pm6\%$ for $n \geq 2$ and $45\%\pm6\%$} for $n \geq 3$) compared to the redder BHB sample (\G{$24\%\pm5\%$ for $n \geq 2$ and $23\%\pm5\%$} for $n \geq 3$), suggesting a potential correlation between binarity and the \G{effective temperature} of BHB stars. 
These observations may help to better constrain the efficiency of envelope stripping during common envelope evolution, offering a modest refinement in our understanding of binary evolution.
We find no significant correlation between $\pi$ ($\kappa$) and metallicity or kinematics, nor with the position of BHB stars on the \G{Kiel diagram}.

\section{Data availability}
Tables 1 is only available in electronic form at the CDS via anonymous ftp to cdsarc.u-strasbg.fr (/CatS/149.217.71.72:001) or via http://cdsweb.u-strasbg.fr/cgi-bin/qcat?J/A+A/.

\begin{acknowledgements}
This work is supported by the Natural Science Foundation of China (Nos.\ 12288102,12125303,12090040/3,12103064,12403039,12373036), the National Key R\&D Program of China (grant Nos. 2021YFA1600403/1, 2021YFA1600400), and the Natural science Foundation of Yunnan Province (Nos. 202201BC070003, 202001AW070007), the International Centre of Supernovae, Yunnan Key Laboratory (No. 202302AN360001), the “Yunnan Revitalization Talent Support Program"-Science, Technology Champion Project (N0. 202305AB350003), and Yunnan Fundamental Research Projects (grant Nos. 202501CF070018).
\end{acknowledgements}

\bibliographystyle{aa} 
\bibliography{aa55002-25} 

\begin{thebibliography}{75}
\expandafter\ifx\csname natexlab\endcsname\relax\def\natexlab#1{#1}\fi

\bibitem[{{Aerts} {et~al.}(2010){Aerts}, {Christensen-Dalsgaard}, \&
  {Kurtz}}]{2010Aerts}
{Aerts}, C., {Christensen-Dalsgaard}, J., \& {Kurtz}, D.~W. 2010,
  {Asteroseismology}

\bibitem[{{Arp} {et~al.}(1952){Arp}, {Baum}, \& {Sandage}}]{1952Arp}
{Arp}, H.~C., {Baum}, W.~A., \& {Sandage}, A.~R. 1952, \aj, 57, 4

\bibitem[{{Bailer-Jones} {et~al.}(2021){Bailer-Jones}, {Rybizki}, {Fouesneau},
  {Demleitner}, \& {Andrae}}]{2021Bailer-Jones}
{Bailer-Jones}, C.~A.~L., {Rybizki}, J., {Fouesneau}, M., {Demleitner}, M., \&
  {Andrae}, R. 2021, \aj, 161, 147

\bibitem[{{Bailyn}(1995)}]{1995Bailyn}
{Bailyn}, C.~D. 1995, \araa, 33, 133

\bibitem[{{Banyard} {et~al.}(2021){Banyard}, {Sana}, {Mahy}, {Bodensteiner},
  {Villase{\~n}or}, \& {Evans}}]{2021Banyard}
{Banyard}, G., {Sana}, H., {Mahy}, L., {et~al.} 2021, arXiv e-prints,
  arXiv:2108.07814

\bibitem[{{Beers} {et~al.}(1996){Beers}, {Wilhelm}, {Doinidis}, \&
  {Mattson}}]{1996Beers}
{Beers}, T.~C., {Wilhelm}, R., {Doinidis}, S.~P., \& {Mattson}, C.~J. 1996,
  \apjs, 103, 433

\bibitem[{{Behr}(2003)}]{2003Behr}
{Behr}, B.~B. 2003, \apjs, 149, 67

\bibitem[{{Belokurov} {et~al.}(2020){Belokurov}, {Sanders}, {Fattahi}, {Smith},
  {Deason}, {Evans}, \& {Grand}}]{2020Belokurov}
{Belokurov}, V., {Sanders}, J.~L., {Fattahi}, A., {et~al.} 2020, \mnras, 494,
  3880

\bibitem[{{Carraro} {et~al.}(1998){Carraro}, {Ng}, \&
  {Portinari}}]{1998Carraro}
{Carraro}, G., {Ng}, Y.~K., \& {Portinari}, L. 1998, \mnras, 296, 1045

\bibitem[{{Castellani} \& {Renzini}(1968)}]{1968Castellani}
{Castellani}, V. \& {Renzini}, A. 1968, \apss, 2, 310

\bibitem[{{Catelan}(2009)}]{2009Catelan}
{Catelan}, M. 2009, \apss, 320, 261

\bibitem[{{Chen} {et~al.}(2018){Chen}, {Li}, \& {Han}}]{2018XueFei}
{Chen}, X., {Li}, Y., \& {Han}, Z. 2018, Scientia Sinica Physica, Mechanica
  \&amp; Astronomica, 48, 079803

\bibitem[{{Chen} {et~al.}(2024){Chen}, {Liu}, \& {Han}}]{2024Chen}
{Chen}, X., {Liu}, Z., \& {Han}, Z. 2024, Progress in Particle and Nuclear
  Physics, 134, 104083

\bibitem[{{Cui} {et~al.}(2012){Cui}, {Zhao}, {Chu}, {Li}, {Li}, {Zhang}, {Su},
  {Yao}, {Wang}, {Xing}, {Li}, {Zhu}, {Wang}, {Gu}, {Luo}, {Xu}, {Zhang},
  {Liu}, {Zhang}, {Yang}, {Cao}, {Chen}, {Chen}, {Chen}, {Chen}, {Chu}, {Feng},
  {Gong}, {Hou}, {Hu}, {Hu}, {Hu}, {Jia}, {Jiang}, {Jiang}, {Jiang}, {Jin},
  {Li}, {Li}, {Li}, {Liu}, {Liu}, {Lu}, {Mao}, {Men}, {Qi}, {Qi}, {Shi},
  {Tang}, {Tao}, {Wang}, {Wang}, {Wang}, {Wang}, {Wang}, {Wang}, {Wang},
  {Wang}, {Wang}, {Wang}, {Wang}, {Wang}, {Xu}, {Xu}, {Yang}, {Yu}, {Yuan},
  {Yuan}, {Zhai}, {Zhang}, {Zhang}, {Zhang}, {Zhao}, {Zhou}, {Zhou}, {Zhu}, \&
  {Zou}}]{2012CuiXiangQun}
{Cui}, X.-Q., {Zhao}, Y.-H., {Chu}, Y.-Q., {et~al.} 2012, Research in Astronomy
  and Astrophysics, 12, 1197

\bibitem[{{Culpan} {et~al.}(2024){Culpan}, {Dorsch}, {Geier}, {Pelisoli},
  {Heber}, {Kub{\'a}tov{\'a}}, \& {Cabezas}}]{2024Culpan}
{Culpan}, R., {Dorsch}, M., {Geier}, S., {et~al.} 2024, \aap, 685, A134

\bibitem[{{Culpan} {et~al.}(2021){Culpan}, {Pelisoli}, \& {Geier}}]{2021Culpan}
{Culpan}, R., {Pelisoli}, I., \& {Geier}, S. 2021, \aap, 654, A107

\bibitem[{{Deng} {et~al.}(2012){Deng}, {Newberg}, {Liu}, {Carlin}, {Beers},
  {Chen}, {Chen}, {Christlieb}, {Grillmair}, {Guhathakurta}, {Han}, {Hou},
  {Lee}, {L{\'e}pine}, {Li}, {Liu}, {Pan}, {Sellwood}, {Wang}, {Wang}, {Yang},
  {Yanny}, {Zhang}, {Zhang}, {Zheng}, \& {Zhu}}]{2012DengLiCai}
{Deng}, L.-C., {Newberg}, H.~J., {Liu}, C., {et~al.} 2012, Research in
  Astronomy and Astrophysics, 12, 735

\bibitem[{{Dorman}(1995)}]{1995Dorman}
{Dorman}, B. 1995, in Liege International Astrophysical Colloquia, Vol.~32,
  Liege International Astrophysical Colloquia, 291

\bibitem[{{Dorman} {et~al.}(1993){Dorman}, {Rood}, \& {O'Connell}}]{1993Dorman}
{Dorman}, B., {Rood}, R.~T., \& {O'Connell}, R.~W. 1993, \apj, 419, 596

\bibitem[{{Duch{\^e}ne} \& {Kraus}(2013)}]{2013Duchene}
{Duch{\^e}ne}, G. \& {Kraus}, A. 2013, \araa, 51, 269

\bibitem[{{Dunstall} {et~al.}(2015){Dunstall}, {Dufton}, {Sana}, {Evans},
  {Howarth}, {Sim{\'o}n-D{\'\i}az}, {de Mink}, {Langer}, {Ma{\'\i}z
  Apell{\'a}niz}, \& {Taylor}}]{2015Dunstall}
{Dunstall}, P.~R., {Dufton}, P.~L., {Sana}, H., {et~al.} 2015, \aap, 580, A93

\bibitem[{{Faulkner}(1966)}]{1966Faulkner}
{Faulkner}, J. 1966, \apj, 144, 978

\bibitem[{{Gaia Collaboration} {et~al.}(2019){Gaia Collaboration}, {Eyer},
  {Rimoldini}, {Audard}, {Anderson}, {Nienartowicz}, {Glass}, {Marchal},
  {Grenon}, {Mowlavi}, {Holl}, {Clementini}, {Aerts}, {Mazeh}, {Evans},
  {Szabados}, {Brown}, {Vallenari}, {Prusti}, {de Bruijne}, {Babusiaux},
  {Bailer-Jones}, {Biermann}, {Jansen}, {Jordi}, {Klioner}, {Lammers},
  {Lindegren}, {Luri}, {Mignard}, {Panem}, {Pourbaix}, {Randich}, {Sartoretti},
  {Siddiqui}, {Soubiran}, {van Leeuwen}, {Walton}, {Arenou}, {Bastian},
  {Cropper}, {Drimmel}, {Katz}, {Lattanzi}, {Bakker}, {Cacciari},
  {Casta{\~n}eda}, {Chaoul}, {Cheek}, {De Angeli}, {Fabricius}, {Guerra},
  {Masana}, {Messineo}, {Panuzzo}, {Portell}, {Riello}, {Seabroke}, {Tanga},
  {Th{\'e}venin}, {Gracia-Abril}, {Comoretto}, {Garcia-Reinaldos}, {Teyssier},
  {Altmann}, {Andrae}, {Bellas-Velidis}, {Benson}, {Berthier}, {Blomme},
  {Burgess}, {Busso}, {Carry}, {Cellino}, {Clotet}, {Creevey}, {Davidson}, {De
  Ridder}, {Delchambre}, {Dell'Oro}, {Ducourant},
  {Fern{\'a}ndez-Hern{\'a}ndez}, {Fouesneau}, {Fr{\'e}mat}, {Galluccio},
  {Garc{\'\i}a-Torres}, {Gonz{\'a}lez-N{\'u}{\~n}ez}, {Gonz{\'a}lez-Vidal},
  {Gosset}, {Guy}, {Halbwachs}, {Hambly}, {Harrison}, {Hern{\'a}ndez},
  {Hestroffer}, {Hodgkin}, {Hutton}, {Jasniewicz}, {Jean-Antoine-Piccolo},
  {Jordan}, {Korn}, {Krone-Martins}, {Lanzafame}, {Lebzelter}, {L{\"o}ffler},
  {Manteiga}, {Marrese}, {Mart{\'\i}n-Fleitas}, {Moitinho}, {Mora}, {Muinonen},
  {Osinde}, {Pancino}, {Pauwels}, {Petit}, {Recio-Blanco}, {Richards}, {Robin},
  {Sarro}, {Siopis}, {Smith}, {Sozzetti}, {S{\"u}veges}, {Torra}, {van Reeven},
  {Abbas}, {Abreu Aramburu}, {Accart}, {Altavilla}, {{\'A}lvarez}, {Alvarez},
  {Alves}, {Andrei}, {Anglada Varela}, {Antiche}, {Antoja}, {Arcay},
  {Astraatmadja}, {Bach}, {Baker}, {Balaguer-N{\'u}{\~n}ez}, {Balm}, {Barache},
  {Barata}, {Barbato}, {Barblan}, {Barklem}, {Barrado}, {Barros}, {Barstow},
  {Bartholom{\'e} Mu{\~n}oz}, {Bassilana}, {Becciani}, {Bellazzini},
  {Berihuete}, {Bertone}, {Bianchi}, {Bienaym{\'e}}, {Blanco-Cuaresma}, {Boch},
  {Boeche}, {Bombrun}, {Borrachero}, {Bossini}, {Bouquillon}, {Bourda},
  {Bragaglia}, {Bramante}, {Breddels}, {Bressan}, {Brouillet},
  {Br{\"u}semeister}, {Brugaletta}, {Bucciarelli}, {Burlacu}, {Busonero},
  {Butkevich}, {Buzzi}, {Caffau}, {Cancelliere}, {Cannizzaro}, {Cantat-Gaudin},
  {Carballo}, {Carlucci}, {Carrasco}, {Casamiquela}, {Castellani},
  {Castro-Ginard}, {Charlot}, {Chemin}, {Chiavassa}, {Cocozza}, {Costigan},
  {Cowell}, {Crifo}, {Crosta}, {Crowley}, {Cuypers}, {Dafonte}, \&
  {Damerdji}}]{2019GaiaCollaboration}
{Gaia Collaboration}, {Eyer}, L., {Rimoldini}, L., {et~al.} 2019, \aap, 623,
  A110

\bibitem[{{Gao} {et~al.}(2017){Gao}, {Zhao}, {Yang}, \& {Gao}}]{2017Gao}
{Gao}, S., {Zhao}, H., {Yang}, H., \& {Gao}, R. 2017, \mnras, 469, L68

\bibitem[{{Gao} {et~al.}(2025){Gao}, {Chen}, {Wang}, \& {Liu}}]{2025GaoXinyi}
{Gao}, X., {Chen}, X., {Wang}, S., \& {Liu}, J. 2025, \apjs, 276, 57

\bibitem[{{Girardi}(2016)}]{2016Girardi}
{Girardi}, L. 2016, \araa, 54, 95

\bibitem[{{Gray} \& {Corbally}(2009)}]{2009Gray}
{Gray}, R.~O. \& {Corbally}, J., C. 2009, {Stellar Spectral Classification}

\bibitem[{{Guo} {et~al.}(2022{\natexlab{a}}){Guo}, {Li}, {Xiong}, {Li}, {Wang},
  {Xiong}, {Luo}, {Hou}, {Liu}, {Han}, \& {Chen}}]{2021GYJfb}
{Guo}, Y., {Li}, J., {Xiong}, J., {et~al.} 2022{\natexlab{a}}, Research in
  Astronomy and Astrophysics, 22, 025009

\bibitem[{{Guo} {et~al.}(2022{\natexlab{b}}){Guo}, {Liu}, {Wang}, {Wang},
  {Zhang}, {Ji}, {Han}, \& {Chen}}]{2022guo886}
{Guo}, Y., {Liu}, C., {Wang}, L., {et~al.} 2022{\natexlab{b}}, \aap, 667, A44

\bibitem[{{Han}(2008)}]{2008Han}
{Han}, Z. 2008, \aap, 484, L31

\bibitem[{{Han} {et~al.}(2003){Han}, {Podsiadlowski}, {Maxted}, \&
  {Marsh}}]{2003Han}
{Han}, Z., {Podsiadlowski}, P., {Maxted}, P.~F.~L., \& {Marsh}, T.~R. 2003,
  \mnras, 341, 669

\bibitem[{{Han} {et~al.}(2002){Han}, {Podsiadlowski}, {Maxted}, {Marsh}, \&
  {Ivanova}}]{2002Han}
{Han}, Z., {Podsiadlowski}, P., {Maxted}, P.~F.~L., {Marsh}, T.~R., \&
  {Ivanova}, N. 2002, \mnras, 336, 449

\bibitem[{{Han} {et~al.}(2020){Han}, {Ge}, {Chen}, \& {Chen}}]{2020Hanzhanwen}
{Han}, Z.-W., {Ge}, H.-W., {Chen}, X.-F., \& {Chen}, H.-L. 2020, Research in
  Astronomy and Astrophysics, 20, 161

\bibitem[{{Heber}(2009)}]{2009Hebersdb}
{Heber}, U. 2009, \araa, 47, 211

\bibitem[{{Heber}(2016)}]{2016Heber}
{Heber}, U. 2016, \pasp, 128, 082001

\bibitem[{{Hoyle} \& {Schwarzschild}(1955)}]{1955Hoyle}
{Hoyle}, F. \& {Schwarzschild}, M. 1955, \apjs, 2, 1

\bibitem[{{Hubrig} {et~al.}(2009){Hubrig}, {Castelli}, {de Silva},
  {Gonz{\'a}lez}, {Momany}, {Netopil}, \& {Moehler}}]{2009Hubrig}
{Hubrig}, S., {Castelli}, F., {de Silva}, G., {et~al.} 2009, \aap, 499, 865

\bibitem[{{Iben} \& {Rood}(1970)}]{1970Iben}
{Iben}, Jr., I. \& {Rood}, R.~T. 1970, \apj, 161, 587

\bibitem[{{Ju} {et~al.}(2024){Ju}, {Cui}, {Huo}, {Liu}, {Xue}, {Liu}, {Feng},
  {Sun}, \& {Li}}]{2024JuBHB}
{Ju}, J., {Cui}, W., {Huo}, Z., {et~al.} 2024, \apjs, 270, 11

\bibitem[{{Kobulnicky} \& {Fryer}(2007)}]{2007Kobulnicky}
{Kobulnicky}, H.~A. \& {Fryer}, C.~L. 2007, \apj, 670, 747

\bibitem[{{Layden} {et~al.}(1996){Layden}, {Hanson}, {Hawley}, {Klemola}, \&
  {Hanley}}]{1996Layden}
{Layden}, A.~C., {Hanson}, R.~B., {Hawley}, S.~L., {Klemola}, A.~R., \&
  {Hanley}, C.~J. 1996, \aj, 112, 2110

\bibitem[{{Lee} {et~al.}(1990){Lee}, {Demarque}, \& {Zinn}}]{1990Lee}
{Lee}, Y.-W., {Demarque}, P., \& {Zinn}, R. 1990, \apj, 350, 155

\bibitem[{{Lee} {et~al.}(1994){Lee}, {Demarque}, \& {Zinn}}]{1994Lee}
{Lee}, Y.-W., {Demarque}, P., \& {Zinn}, R. 1994, \apj, 423, 248

\bibitem[{{Li} {et~al.}(2024){Li}, {Zhang}, {Chen}, {Ge}, {Jiang}, {Li},
  {Chen}, \& {Han}}]{2024LiZW}
{Li}, Z., {Zhang}, Y., {Chen}, H., {et~al.} 2024, \apj, 964, 22

\bibitem[{{Liebert} {et~al.}(1994){Liebert}, {Saffer}, \&
  {Green}}]{1994Liebert}
{Liebert}, J., {Saffer}, R.~A., \& {Green}, E.~M. 1994, \aj, 107, 1408

\bibitem[{{Liu}(2019)}]{2019liuchaosmokinggun}
{Liu}, C. 2019, \mnras, 490, 550

\bibitem[{{Liu} {et~al.}(2020){Liu}, {Fu}, {Shi}, {Wu}, {Han}, {Chen}, {Dong},
  {Zhao}, {Chen}, {Zhang}, {Bai}, {Chen}, {Cui}, {Du}, {Hsia}, {Jiang}, {Hou},
  {Hou}, {Li}, {Li}, {Li}, {Liu}, {Liu}, {Luo}, {Ren}, {Tian}, {Tian}, {Wang},
  {Wu}, {Xie}, {Yan}, {Yang}, {Yu}, {Zhang}, {Zhang}, {Zhang}, {Zhang}, {Zhao},
  {Zhong}, {Zong}, \& {Zuo}}]{2020LiuChao}
{Liu}, C., {Fu}, J., {Shi}, J., {et~al.} 2020, arXiv e-prints, arXiv:2005.07210

\bibitem[{{Luo} {et~al.}(2021){Luo}, {Zhao}, {Li}, {Guo}, \&
  {Liu}}]{2021luofeng}
{Luo}, F., {Zhao}, Y.-H., {Li}, J., {Guo}, Y.-J., \& {Liu}, C. 2021, Research
  in Astronomy and Astrophysics, 21, 272

\bibitem[{{Mahy} {et~al.}(2021){Mahy}, {Lanthermann}, {Hutsem{\'e}kers},
  {Kluska}, {Lobel}, {Manick}, {Miszalski}, {Reggiani}, {Sana}, \&
  {Gosset}}]{2021MahyC}
{Mahy}, L., {Lanthermann}, C., {Hutsem{\'e}kers}, D., {et~al.} 2021, arXiv
  e-prints, arXiv:2105.12380

\bibitem[{{Maxted} {et~al.}(2001){Maxted}, {Heber}, {Marsh}, \&
  {North}}]{2001Maxted}
{Maxted}, P.~F.~L., {Heber}, U., {Marsh}, T.~R., \& {North}, R.~C. 2001,
  \mnras, 326, 1391

\bibitem[{{Moe} \& {Di Stefano}(2017)}]{2017MoeStefano}
{Moe}, M. \& {Di Stefano}, R. 2017, \apjs, 230, 15

\bibitem[{{Moe} {et~al.}(2019){Moe}, {Kratter}, \& {Badenes}}]{2019Moe}
{Moe}, M., {Kratter}, K.~M., \& {Badenes}, C. 2019, \apj, 875, 61

\bibitem[{{Moehler}(2001)}]{2001Moehler}
{Moehler}, S. 2001, \pasp, 113, 1162

\bibitem[{{Moehler}(2025)}]{2025Moehler}
{Moehler}, S. 2025, \aap, 693, A136

\bibitem[{{{\O}stensen} {et~al.}(2012){{\O}stensen}, {Degroote}, {Telting},
  {Vos}, {Aerts}, {Jeffery}, {Green}, {Reed}, \& {Heber}}]{2012Ostensen}
{{\O}stensen}, R.~H., {Degroote}, P., {Telting}, J.~H., {et~al.} 2012, \apjl,
  753, L17

\bibitem[{{Pier}(1984)}]{1984Pier}
{Pier}, J.~R. 1984, \apj, 281, 260

\bibitem[{{Raghavan} {et~al.}(2010){Raghavan}, {McAlister}, {Henry}, {Latham},
  {Marcy}, {Mason}, {Gies}, {White}, \& {ten Brummelaar}}]{2010Raghavan}
{Raghavan}, D., {McAlister}, H.~A., {Henry}, T.~J., {et~al.} 2010, \apjs, 190,
  1

\bibitem[{{Rich} {et~al.}(1997){Rich}, {Sosin}, {Djorgovski}, {Piotto}, {King},
  {Renzini}, {Phinney}, {Dorman}, {Liebert}, \& {Meylan}}]{1997Rich}
{Rich}, R.~M., {Sosin}, C., {Djorgovski}, S.~G., {et~al.} 1997, \apjl, 484, L25

\bibitem[{{Rocha-Pinto} {et~al.}(2000){Rocha-Pinto}, {Maciel}, {Scalo}, \&
  {Flynn}}]{2000Rocha-Pinto}
{Rocha-Pinto}, H.~J., {Maciel}, W.~J., {Scalo}, J., \& {Flynn}, C. 2000, \aap,
  358, 850

\bibitem[{{Ruhland} {et~al.}(2011){Ruhland}, {Bell}, {Rix}, \&
  {Xue}}]{2011Ruhland}
{Ruhland}, C., {Bell}, E.~F., {Rix}, H.-W., \& {Xue}, X.-X. 2011, \apj, 731,
  119

\bibitem[{{Salgado} {et~al.}(2013){Salgado}, {Moni Bidin}, {Villanova},
  {Geisler}, \& {Catelan}}]{2013Salgado}
{Salgado}, C., {Moni Bidin}, C., {Villanova}, S., {Geisler}, D., \& {Catelan},
  M. 2013, \aap, 559, A101

\bibitem[{{Sana} {et~al.}(2013){Sana}, {de Koter}, {de Mink}, {Dunstall},
  {Evans}, {H{\'e}nault-Brunet}, {Ma{\'\i}z Apell{\'a}niz},
  {Ram{\'\i}rez-Agudelo}, {Taylor}, {Walborn}, {Clark}, {Crowther}, {Herrero},
  {Gieles}, {Langer}, {Lennon}, \& {Vink}}]{2013sana}
{Sana}, H., {de Koter}, A., {de Mink}, S.~E., {et~al.} 2013, \aap, 550, A107

\bibitem[{{Sana} {et~al.}(2012){Sana}, {de Mink}, {de Koter}, {Langer},
  {Evans}, {Gieles}, {Gosset}, {Izzard}, {Le Bouquin}, \&
  {Schneider}}]{2012SanaScience}
{Sana}, H., {de Mink}, S.~E., {de Koter}, A., {et~al.} 2012, Science, 337, 444

\bibitem[{{Sana} \& {Evans}(2011)}]{2011Sana}
{Sana}, H. \& {Evans}, C.~J. 2011, in IAU Symposium, Vol. 272, Active OB Stars:
  Structure, Evolution, Mass Loss, and Critical Limits, ed. C.~{Neiner},
  G.~{Wade}, G.~{Meynet}, \& G.~{Peters}, 474--485

\bibitem[{{Searle} \& {Rodgers}(1966)}]{1966Searle}
{Searle}, L. \& {Rodgers}, A.~W. 1966, \apj, 143, 809

\bibitem[{{Sweigart}(1990)}]{1990Sweigart}
{Sweigart}, A.~V. 1990, in Astronomical Society of the Pacific Conference
  Series, Vol.~11, Confrontation Between Stellar Pulsation and Evolution, ed.
  C.~{Cacciari} \& G.~{Clementini}, 1--10

\bibitem[{{Sweigart} \& {Gross}(1976)}]{1976Sweigart}
{Sweigart}, A.~V. \& {Gross}, P.~G. 1976, \apjs, 32, 367

\bibitem[{{Tian} {et~al.}(2024){Tian}, {Liu}, {Li}, \& {Zhang}}]{2024tianhao}
{Tian}, H., {Liu}, C., {Li}, J., \& {Zhang}, B. 2024, \mnras, 531, 1730

\bibitem[{{Tian} {et~al.}(2019){Tian}, {Liu}, {Xu}, \& {Xue}}]{2019tianhao}
{Tian}, H., {Liu}, C., {Xu}, Y., \& {Xue}, X. 2019, \apj, 871, 184

\bibitem[{{Twarog}(1980)}]{1980Twarog}
{Twarog}, B.~A. 1980, \apj, 242, 242

\bibitem[{{Xue} {et~al.}(2011){Xue}, {Rix}, {Yanny}, {Beers}, {Bell}, {Zhao},
  {Bullock}, {Johnston}, {Morrison}, {Rockosi}, {Koposov}, {Kang}, {Liu},
  {Luo}, {Lee}, \& {Weaver}}]{2011Xue}
{Xue}, X.-X., {Rix}, H.-W., {Yanny}, B., {et~al.} 2011, \apj, 738, 79

\bibitem[{{Xue} {et~al.}(2008){Xue}, {Rix}, {Zhao}, {Re Fiorentin}, {Naab},
  {Steinmetz}, {van den Bosch}, {Beers}, {Lee}, {Bell}, {Rockosi}, {Yanny},
  {Newberg}, {Wilhelm}, {Kang}, {Smith}, \& {Schneider}}]{2008Xue}
{Xue}, X.~X., {Rix}, H.~W., {Zhao}, G., {et~al.} 2008, \apj, 684, 1143

\bibitem[{{York} {et~al.}(2000){York}, {Adelman}, {Anderson}, {Anderson},
  {Annis}, {Bahcall}, {Bakken}, {Barkhouser}, {Bastian}, {Berman}, {Boroski},
  {Bracker}, {Briegel}, {Briggs}, {Brinkmann}, {Brunner}, {Burles}, {Carey},
  {Carr}, {Castander}, {Chen}, {Colestock}, {Connolly}, {Crocker}, {Csabai},
  {Czarapata}, {Davis}, {Doi}, {Dombeck}, {Eisenstein}, {Ellman}, {Elms},
  {Evans}, {Fan}, {Federwitz}, {Fiscelli}, {Friedman}, {Frieman}, {Fukugita},
  {Gillespie}, {Gunn}, {Gurbani}, {de Haas}, {Haldeman}, {Harris}, {Hayes},
  {Heckman}, {Hennessy}, {Hindsley}, {Holm}, {Holmgren}, {Huang}, {Hull},
  {Husby}, {Ichikawa}, {Ichikawa}, {Ivezi{\'c}}, {Kent}, {Kim}, {Kinney},
  {Klaene}, {Kleinman}, {Kleinman}, {Knapp}, {Korienek}, {Kron}, {Kunszt},
  {Lamb}, {Lee}, {Leger}, {Limmongkol}, {Lindenmeyer}, {Long}, {Loomis},
  {Loveday}, {Lucinio}, {Lupton}, {MacKinnon}, {Mannery}, {Mantsch}, {Margon},
  {McGehee}, {McKay}, {Meiksin}, {Merelli}, {Monet}, {Munn}, {Narayanan},
  {Nash}, {Neilsen}, {Neswold}, {Newberg}, {Nichol}, {Nicinski}, {Nonino},
  {Okada}, {Okamura}, {Ostriker}, {Owen}, {Pauls}, {Peoples}, {Peterson},
  {Petravick}, {Pier}, {Pope}, {Pordes}, {Prosapio}, {Rechenmacher}, {Quinn},
  {Richards}, {Richmond}, {Rivetta}, {Rockosi}, {Ruthmansdorfer}, {Sandford},
  {Schlegel}, {Schneider}, {Sekiguchi}, {Sergey}, {Shimasaku}, {Siegmund},
  {Smee}, {Smith}, {Snedden}, {Stone}, {Stoughton}, {Strauss}, {Stubbs},
  {SubbaRao}, {Szalay}, {Szapudi}, {Szokoly}, {Thakar}, {Tremonti}, {Tucker},
  {Uomoto}, {Vanden Berk}, {Vogeley}, {Waddell}, {Wang}, {Watanabe},
  {Weinberg}, {Yanny}, {Yasuda}, \& {SDSS Collaboration}}]{2000Yook}
{York}, D.~G., {Adelman}, J., {Anderson}, Jr., J.~E., {et~al.} 2000, \aj, 120,
  1579

\bibitem[{{Zhang} {et~al.}(2021){Zhang}, {Li}, {Yang}, {Xiong}, {Fu}, {Liu},
  {Tian}, {Li}, {Wang}, {Liang}, {Zhou}, {Zong}, {Yang}, {Liu}, \&
  {Hou}}]{2021zhangboRV}
{Zhang}, B., {Li}, J., {Yang}, F., {et~al.} 2021, \apjs, 256, 14

\bibitem[{{Zhao} {et~al.}(2012){Zhao}, {Zhao}, {Chu}, {Jing}, \&
  {Deng}}]{2012ZhaoGang}
{Zhao}, G., {Zhao}, Y.-H., {Chu}, Y.-Q., {Jing}, Y.-P., \& {Deng}, L.-C. 2012,
  Research in Astronomy and Astrophysics, 12, 723

\end{thebibliography}
\end{document}